\documentclass{amsart}

\usepackage{amsmath,amssymb}
\usepackage{graphicx}

\numberwithin{equation}{section}

\def\res{\mathrm {res}}

\def\al{\mathrm {al}}

\def\Sym{\mathrm {Sym}}

\def\CC{{\mathbb C}}

\def\Qp{{\mathrm{Q}}}

\def\Pp{{\mathrm{P}}}
\def\Qp{{\mathrm{Q}}}

\newtheorem{theorem}{Theorem}[section]
\newtheorem{lemma}[theorem]{Lemma}
\newtheorem{corollary}[theorem]{Corollary}

\theoremstyle{definition}
\newtheorem{definition}[theorem]{Definition}

\theoremstyle{remark}

\def\book#1{\rm{#1}, }
\def\paper#1{\textit{#1}, }
\def\jour#1{\rm{#1}, }
\def\yr#1{({\rm{#1}) }}
\def\vol#1{\textbf{#1}}
\def\pages#1{\rm{#1}}

\def\publaddr#1{\rm{#1}, }
\def\publ#1{\rm{#1}, }
\def\by#1{{\rm{#1}, }}

\begin{document}

\title{Relations of al Functions
over Subvarieties in a Hyperelliptic Jacobian}

\author{Shigeki MATSUTANI}
\address{8-21-1 Higashi-Linkan, Sagamihara, 228-0811, JAPAN}

\email{rxb01142@nifty.com}



\maketitle

\begin{abstract}
The sine-Gordon equation has hyperelliptic al function
solutions over a  hyperelliptic Jacobian
for $y^2 = f(x)$ of arbitrary genus $g$.
This article gives an extension of
the sine-Gordon equation to that
over subvarieties of the  hyperelliptic Jacobian.
 We also obtain
the condition that the sine-Gordon equation in
a proper subvariety of the Jacobian is defined.
\end{abstract}

\section{Introduction}

For a hyperelliptic curve $C_g$ given by
an affine curve $y^2 = \prod_{i=1}^{2g+1}(x-b_i)$,
where $b_i$'s are complex numbers,
we have a Jacobian $\mathcal J_g$ as a complex
torus $\CC^g/\Lambda$ by the Abel map $\omega$ \cite{Mu}.
Due to the Abelian theorem, we have a natural
morphism from the symmetrical product $\Sym^g(C_g)$
to the Jacobian $\mathcal J_g\approx\omega[\Sym^{g}(C_g)]/\Lambda$.
As zeros of an appropriate shifted Riemann theta
function over $\mathcal J_g$, the theta divisor
is defined as
$$
	\Theta := \omega[\Sym^{g-1}(C_g)]/\Lambda
$$
which is a subvariety of $\mathcal J_g$.
Similarly, it is natural to introduce a subvariety
$$
	\Theta_k := \omega[\Sym^{k}(C_g)]/\Lambda
$$
and a sequence,
$$
    \Theta_0\subset\Theta_1 \subset \Theta_2 \subset \cdots \subset
\Theta_{g-1}
        \subset \Theta_g \equiv \mathcal J_g
$$
Vanhaecke studied the structure of these subvarieties
as stratifications of the Jacobian $\mathcal J_g$ using the strategies
developed in the studies of the infinite dimensional
integrable system \cite{V}.
He showed that these stratifications of the Jacobian
are connected  with  stratifications of the
Sato Grassmannian.
Further
Vanhaecke investigated Lie-Poisson structures in the
Jacobian in \cite{V2}. He showed that  invariant
manifolds associated with Poisson brackets
can be identified with
these strata; it implies that the strata are
characterized by the Lie-Poisson structures.
He also showed that the Poisson brackets are connected with
a finite-dimensional integrable system,
Henon-Heiles system. Following the study,
Abenda and Fedorov \cite{AF}
investigated these strata and their relations
 to Henon-Heiles system and Neumann systems.

On the other hand, functions over
the embedded hyperelliptic curve $\Theta_1$
in a hyperelliptic Jacobian $\mathcal J_g$
were also studied from viewpoint of number theory in \cite{C, G, O}.
In \cite{O}, \^Onishi also investigated
 the sequence of the subvarieties,
and explicitly studied behaviors of functions over them
in order to obtain higher genus analog of the
 Frobenius-Stickelberger relations for genus one case.
Though Vanhaecke, Abenda and Fedorov
found some relations of functions
over these subvarieties
explicitly using the infinite universal grassmannians
and so-called Mumford's $UVW$ expressions \cite{Mu},
\^Onishi gave more explicit relations on some
functions over the subvarieties
using the theory of hyperelliptic
functions in the nineteenth century fashion
\cite{Ba1, Ba2, Ba3}.

In this article, we will also investigate some relations of
functions over the subvarieties
based upon the studies of the hyperelliptic
function theory developed in the nineteenth century
\cite{Ba2, Ba3, W}.
Especially this article deals with
 the {\lq\lq}sine-Gordon equation{\rq\rq} over there.

Modern expressions of the sine-Gordon equation
 in terms of Riemann theta functions
were given in [\cite{Mu} 3.241],
\begin{gather}
        \frac{\partial}{\partial t_\Pp}
        \frac{\partial}{\partial t_\Qp}
        \log ( [2\Pp-2\Qp] ) =
        A( [2\Pp-2\Qp ] - [2\Qp-2\Pp ])   , \label{eq:1-5}
\end{gather}
where $\Pp$ and $\Qp$ are ramified points of $C_g$,
$A$ is a constant number,
 $[D]$ is a meromorphic function over $\Sym^g(C_g)$
with a divisor $D$ for each $C_g$
and $t_{\Pp'}$ is a coordinate in the Jacobi variety such that
it is identified with a local parameter at a ramified point
$\Pp'$ up to constant.

In the previous work \cite{Ma}, we also studied (\ref{eq:1-5})
 using the fashion of the nineteenth century.
 In \cite{W}
Weierstrass defined $\al$ function  by $\al_r:=\gamma_r
\sqrt{F_g(b_r)}$ and $F_g(z):=(x_1 - z) \cdots (x_g- z)$
over $\mathcal J_g$ with
a constant factor $\gamma_r$.
Let $\gamma_r=1$ in this article.
As Weierstrass implicitly seemed to deal with it,
(\ref{eq:1-5}) is naturally described by al-functions
as \cite{Ma},
\begin{gather}
\frac{\partial^2}
{\partial v^{(g)}_1\partial v_2^{(g)}} \log \frac{\al_r^{}}
{\al_s^{}}
   = \frac{1}{(b_r-b_s)}
\left( f'(b_s)\left(\frac{\al_r^{}}{\al_s^{}}\right)^2 +
       f'(b_r)\left(\frac{\al_s^{}}{\al_r^{}}\right)^2
          \right). \label{eq:1-4}
\end{gather}
Here $f'(x):=d f(x)/dx$ and
  $v^{(g)}$'s are defined in (\ref{eq:2-4}).
((\ref{eq:1-4}) was obtained in the previous article \cite{Ma} by more
direct computations
and will be shown as Corollary 3.3 in this article).
We call (\ref{eq:1-4}) Weierstrass relation in this article.

In this article, we will introduce an
{\lq\lq}al{\rq\rq} function
over the subvariety in the Jacobian,
 $\al_r^{(m)}:=\sqrt{F_m(b_r)}$
and $F_m(z):=(x_1 - z) \cdots (x_m- z)$ for a point
$((x_1,y_1),$ $\cdots,$ $(x_m,y_m))$ in
the symmetric product of the $m$ curves $\Sym^m C_g$
$(m=1,\cdots,g-1)$.
In \cite{Mu}, Mumford dealt with $F_m$ function
(he denoted it by $U$) for $1\le m <g$ and studied the properties.
Further Abenda and Fedorov
also studied some properties of the $\al_r^{(m)}$
and $F_m$ functions in \cite{AF} though
they did not mention about Weierstrass's paper
nor the relation (\ref{eq:1-4}).
We will consider a  variant of  the Weierstrass relation
(\ref{eq:1-4})  to $\al_r^{(m)}$
over subvariety in  non-degenerated and degenerated
hyperelliptic Jacobian.

As in our main theorem 3.1,
even on the subvarieties, we have a similar relation to (\ref{eq:1-5}),
\begin{gather}
\split
\frac{\partial}{\partial v_{r}^{(m)} }
          \frac{\partial}{\partial v_s^{(m)}}
      \log \frac{\al_r^{(m)}}{\al_s^{(m)}}
   &= \frac{1}{(b_r-b_s)}
  \left(\frac{f'(b_r)}{(Q^{(2)}_m(b_r))^2}
\left(\frac{\al_s^{(m)}}{\al_r^{(m)}}\right)^2
+\frac{f'(b_s)}{(Q^{(2)}_m(b_s))^2}
\left(\frac{\al_r^{(m)}}{\al_s^{(m)}}\right)^2\right)\\
   &\quad + \text{reminder terms}.
\endsplit\label{eq:1-6}
\end{gather}
Here $Q^{(2)}_m$ is defined in (\ref{eq:2-2}).
We regard (\ref{eq:1-6}) or (\ref{eq:3-1}) as a
subvariety version of the Weierstrass relation (\ref{eq:1-4}). In fact,
(\ref{eq:1-6}) contains the same form as (\ref{eq:1-5}) up to the factors
$(Q^{(2)}_m(b_t))^2$ ($t=r,s$) and the reminder terms.
Thus (\ref{eq:1-6}) or (\ref{eq:3-1}) should be regarded as an extension
of the sine-Gordon equation (\ref{eq:1-4})
 over the Jacobian
to that over
the subvariety of the Jacobian.

Further a certain degenerate curve, the remainders in (\ref{eq:1-6}) vanishes.
Then we have a relations
 over subvarieties in
the Jacobian, which is formally the same as
the Weierstrass relations (\ref{eq:1-4}) up to the factors
$(Q^{(2)}_m(b_t))^2$ ($t=r,s$),
which means that we can find  solutions of sine-Gordon
equation  over subvarieties in hyperelliptic Jacobian.
We expect that our results shed a light on the new field
of a relation between {\lq\lq}integrability{\rq\rq} and
a subvariety
in the Jacobian, which was brought off by
\cite{V, V2, AF}.

\vskip 0.5 cm
The author is grateful to the referee for directing his
attensions to the references \cite{AF} and \cite{V2}.

\vskip 0.5 cm

\section{Differentials of a Hyperelliptic Curve}

\vskip 0.5 cm

In this section, we will give our conventions  of
 hyperelliptic functions
of  a hyperelliptic curve $C_g$  of genus $g$
$(g>0)$ given by an affine equation,
\begin{gather}
 \split
   y^2 &= f(x)= (x-b_1)(x-b_2)\cdots(x-b_{2g})(x-b_{2g+1})\\
             &= Q(x)P(x),\\
\endsplit  \label{eq:2-1}
\end{gather}
where $b_j$'s are  complex numbers.
Here we use the expressions  $Q(x):= Q_m^{(1)}(x)Q_m^{(2)}(x)$,
\begin{gather}
\split
        Q_m^{(1)}(x) &:= (x-a_1)(x-a_2)\cdots(x-a_{m}),\\
        Q_m^{(2)}(x) &:= (x-a_{m+1})(x-a_{m+2})\cdots(x-a_{g}),\\
        P(x) &:= (x-c_1)(x-c_2)
                   \cdots(x-c_{g})(x-c),\\
\endsplit \label{eq:2-2}
\end{gather}
where $a_k\equiv b_k$, $c_k \equiv b_{g+k}$, $(k=1,\cdots,g)$
$c\equiv b_{2g+1}$.

\begin{definition}\cite{Ba1, Ba2}

 \begin{enumerate}
For a point $(x_i, y_i)\in C_g$, we define the following
quantities.

\item The unnormalized differentials of the first kind are
defined by,
\begin{gather}
   d v^{(g,i)}_k := \frac{Q(x_i) d x_i}{2(x_i-a_k)Q'(a_k)y_i},
        \quad(k=1,\cdots,g)
      \label{eq:2-3}
\end{gather}

\item
The Abel map for $g$-th symmetric product
of the curve $C_g$ is defined by,
\begin{gather*}
 v^{(g)}\equiv(v_1^{(g)},\cdots,v_g^{(g)})
:\Sym^g( C_g) \longrightarrow \Bbb C^g,
\end{gather*}
\begin{gather}
      \left( v_k^{(g)}((x_1,y_1),\cdots,(x_g,y_g)):= \sum_{i=1}^g
       \int_{\infty}^{(x_i,y_i)} d v^{(g,i)}_k \right).
      \label{eq:2-4}
\end{gather}

\item
For $v^{(g)}\in \CC^g$, we define the subspace,
\begin{gather}
  \Xi_m  :=  v^{(g)}(\Sym^m( C_g)
              \times(a_{m+1},0)\times\cdots\times(a_{g},0)) /{ \pmb{\Lambda}} .
     \label{eq:2-5}
\end{gather}
Here  $\CC$ is a complex field and
${ \pmb{\Lambda}}$  is a $g$-dimensional
lattice generated by the related periods
or the hyperelliptic integrals of
the first kind.

\end{enumerate}
\end{definition}

The Jacobi variety  $\mathcal J_g$
are defined as complex torus as $\mathcal J_g := \Xi_g$.
As  $\Xi_m$ ($m<g$) is embedded in $\mathcal J_g$  whose complex dimension as
subvariety is $m$,
the differential forms $(d v_k^{(g)})_{k=1,\cdots,g}$ are not
linearly independent. We select linearly independent bases such as
$v_k^{(m)}:=v_k^{(g)}((x_1,y_1),\cdots,(x_m,y_m),
(a_{m+1},0),\cdots,(a_{g},0))$, $(k=1,\cdots,m)$ at $\Xi_m$.
$$
\Xi_0 \subset \Xi_1 \subset \Xi_2 \subset \cdots \subset \Xi_{g-1}
        \subset \Xi_g \equiv J_g
$$

\bigskip
For $(x_1,\cdots,x_m) \in \Sym^m( C_g)$, we introduce
\begin{gather}
	F_m(x):= (x-x_1) \cdots (x-x_m),
          \label{eq:2-6}
\end{gather}
 and in terms of $F_m(x)$, a
hyperelliptic $al$-function over $(v^{(m)})\in \Xi_m$,
[Ba2 p.340, W],
\begin{gather}
\al_r^{(m)}(v^{(m)}) =\sqrt{F_m(b_r)} . \label{eq:2-7}
\end{gather}
Further we introduce $m\times m$-matrices,
\begin{gather*}
\mathcal M_m := \begin{pmatrix}
     \dfrac{1}{x_1-a_1} &  \dfrac{1}{x_2-a_1} & \cdots & \dfrac{1}{x_m-a_1} \\
     \dfrac{1}{x_1-a_2} &  \dfrac{1}{x_2-a_2} & \cdots & \dfrac{1}{x_m-a_2} \\
   \vdots & \vdots & \ddots & \vdots  \\
     \dfrac{1}{x_1-a_m} &  \dfrac{1}{x_2-a_m} & \cdots & \dfrac{1}{x_m-a_m}
     \end{pmatrix},
\end{gather*}
\begin{gather*}
	\mathcal Q_m = \begin{pmatrix}
     \sqrt{ \dfrac{Q(x_1)}{P(x_1)}} & \ & \ & \  \\
      \ &\sqrt{ \dfrac{Q(x_2)}{P(x_2)}}& \ & \   \\
      \ & \ & \ddots   & \   \\
      \ & \ & \ & \sqrt{ \dfrac{Q(x_m)}{P(x_m)}}  \end{pmatrix},
\quad
	\mathcal A_m = \begin{pmatrix}
      Q'(a_1) & \ & \ & \  \\
      \ &  Q'(a_2)& \ & \   \\
      \ & \ & \ddots   & \   \\
      \ & \ & \ &   Q'(a_m) \end{pmatrix}.
\end{gather*}

\begin{lemma}

\begin{enumerate}

\item
\begin{gather*}
\det\mathcal M_m = \frac{(-1)^{m(m-1)/2} P(x_1,\cdots,x_m)P(a_1,\cdots,a_m)}
            {\prod_{k,l}(x_k-a_l)},
\end{gather*}
where
\begin{gather*}
        P(z_1,\cdots,z_m)  :=\prod_{i<j} (z_i - z_j ).
\end{gather*}

\item
\begin{gather*}
 \mathcal M_m^{-1} =\left[\left( \frac{F_m(a_j)Q_m^{(1)}(x_i)}
     {F_m'(x_i)Q_m^{(1)\prime}(a_j)(a_j-x_i)}
                     \right)_{i,j} \right],
\end{gather*}
where $F_m'(x):=d F_m(x)/d x$ and $Q_m^{(1)\prime}(x)
=d Q_m^{(1)}(x)/dx$.

\item
\begin{gather}
( \mathcal M \mathcal Q)^{-1}\mathcal A=
        \left[\left( \frac{2 y_i F_m(a_j)}{F_m'(x_i)Q_m^{(2)}(x_i)(a_j-x_i)}
                     \right)_{i,j} \right].\label{eq:2-8}
\end{gather}

\end{enumerate}
\end{lemma}

\begin{proof}
(1) is a well-known result \cite{T}. Since the zero
and singularity in the left hand side give the right hand side as
$$
C \Pp(x_1,\cdots,x_m)\Pp(a_1,\cdots,a_m)/{\prod_{k,l}(x_k-a_l)},
$$
for a certain constant $C$. In order to determine $C$, we multiply
${\prod_{k,l}(x_k-a_l)}$ both sides and let $x_1=a_1$,
$x_2=a_2$, $\cdots$, and $x_m=a_m$. Then $C$ is determined as above.
(2) is obtained by the Laplace formula using the minor determinant
for the inverse matrix. On (3) we note that $Q_m^{(1)}Q_m^{(2)}=Q(x)$
in (\ref{eq:2-2})
and thus $Q_m^{(1)}(x)\sqrt{P(x)/Q(x)} =y / Q_m^{(2)}$. Then we obtain (3).
\end{proof}

\begin{corollary}
Let
$\partial_{v_i}^{(r)}:=\partial/\partial{v_i^{(r)}}$,
and
$\partial_{x_i}:=\partial/\partial{x_i}$.
\begin{gather}
	\begin{pmatrix} \partial_{v_1}\\
                 \partial_{v_2}\\
                 \vdots\\
                 \partial_{v_m}
         \end{pmatrix}
   =2( \mathcal M \mathcal Q_m)^{-1}\mathcal A_m
	\begin{pmatrix} \partial_{x_1}\\
                 \partial_{x_2}\\
                 \vdots\\
                 \partial_{x_m}
         \end{pmatrix}. \label{eq:2-9}
\end{gather}
\end{corollary}

\section{Weierstrass relation on $\Xi_m$}

The hyperelliptic solution of the sine-Gordon
equation over $\mathcal J_g$ related to ramified points $(a_1,0)$ and
$(a_2,0)$ is obtained as (\ref{eq:1-5}) by Mumford \cite{Mu},
whose expression in an old fashion is the Weierstrass relation (\ref{eq:1-4}).
Let us consider an extension of the Weierstrass relation (\ref{eq:1-4})
 over $\Xi_m$ as our main theorem.
We will give the theorem as follows.

\begin{theorem}
$\al_r^{(m)}$ and  $\al_s^{(m)}$ $(r, s \in \{1,2,\cdots,m\})$
over $\Xi_m$ in (\ref{eq:2-5}) obey the relation,
\begin{gather}
\split
\frac{\partial}{\partial v_{r}^{(m)} }
          \frac{\partial}{\partial v_s^{(m)}}
     & \log \frac{\al_r^{(m)}(v^{(m)})}{\al_s^{(m)}(v^{(m)})}\\
   &= \frac{1}{(a_r-a_s)}
  \left(\frac{f'(a_r)}{(Q^{(2)}_m(a_r))^2}
       \left(\frac{\al_s^{(m)}(v^{(m)})}{\al_r^{(m)}(v^{(m)})}\right)^2
+\frac{f'(a_s)}{(Q^{(2)}_m(a_s))^2}
 \left(\frac{\al_r^{(m)}(v^{(m)})}{\al_s^{(m)}(v^{(m)})}\right)^2\right)\\
&+\frac{f'(a_{m+1})(\al_r^{(m)}(v^{(m)}))^2(\al_s^{(m)}(v^{(m)}))^2(a_r-a_s)}
{(a_{m+1}-a_r)(a_{m+1}-a_s)(\al_{m+1}^{(m)}(v^{(m)}))^4
(Q_m^{(2)\prime}(a_{m+1}))^2}\\
&+\cdots \cdots\\
&+\frac{f'(a_{g})(\al_r^{(m)}(v^{(m)}))^2(\al_s^{(m)}(v^{(m)}))^2(a_r-a_s)}
{(a_{g}-a_r)(a_{g}-a_s)(\al_{g}^{(m)}(v^{(m)}))^4(Q_m^{(2)}(a_g)')^2}.\\
\endsplit
      \label{eq:3-1}
\end{gather}

\end{theorem}

\begin{proof}
From (\ref{eq:2-7}), we will consider the following formula instead of
(\ref{eq:3-1}) without loss of generality,
\begin{gather}
\split
\frac{\partial}{\partial v_{1}^{(m)} }
          \frac{\partial}{\partial v_2^{(m)}}\log \frac{F_m(a_1)}{F_m(a_2)}
   &=2 \frac{F_m(a_1)F_m(a_2)}{(a_1-a_2)}
  \Bigr(\frac{f'(a_1)}{F_m(a_1)^2(Q^{(2)}_m(a_1))^2}
+\frac{f'(a_2)}{F_m(a_2)^2(Q^{(2)}_m(a_1))^2}\\
&+\frac{f'(a_{m+1})(a_1-a_2)^2}{
(a_{m+1}-a_1)(a_{m+1}-a_2)F_m(a_{m+1})^2 (Q^{(2)\prime}_m(a_{m+1}))^2}\\
&+\cdots \\
&+\frac{f'(a_{g})(a_1-a_2)^2}{
(a_g-a_1)(a_g-a_2)F_m(a_{g})^2 (Q^{(2)\prime}_m(a_g))^2}  \Bigr).\\
\endsplit
      \label{eq:3-2}
\end{gather}
Before we start the proof, we will comment on our strategy,
which is essentially the same as \cite{Ba3}.
First we translate the words of the Jacobian into those of
the curves; we rewrite the differentials $v_{(r)}^{(m)}$'s
in terms of the differentials over curves as in (\ref{eq:3-3}).
We count the
residue of an integration and use a combinatorial trick as
in Lemma 3.2.
Then we will obtain (\ref{eq:3-2}).

From (\ref{eq:2-8}) and
 (\ref{eq:2-9}), the derivative $v$'s over $\Xi_m$ in (\ref{eq:2-5})
are expressed
 by the affine coordinate $x_i$'s,
\begin{gather}
	\frac{\partial}{\partial v_i^{(m)} }=F_m(a_i)Q^{(2)}_m(a_i)
         \sum_{j=1}^m \frac{2y_j}{F_m'(x_j)Q^{(2)}_m(x_j)(x_j-a_i)
                   }
     \frac{\partial}{\partial x_j}.\label{eq:3-3}
\end{gather}
The right hand side of (\ref{eq:3-2}) becomes,
\begin{gather*}
\split
&\frac{\partial^2}{\partial v_1\partial v_2 }\log \frac{F_m(a_1)}{F_m(a_2)}
   =F_m(a_1)Q^{(2)}_m(a_1)\\
  &
 \sum_{j=1i=1}^m\frac{2y_j}{( x_i-a_1)F_m'(x_j)Q^{(2)}_m(x_j)}
\frac{\partial}{\partial x_j}
         \frac{2y_iF_m(a_2)Q^{(2)}_m(a_1)}{F_m'(x_i) Q^{(2)}_m(x_i)(x_i-a_2)}
         \frac{(a_1-a_2)}{(x_i-a_1)(x_i-a_2)}.\\
\endsplit
\end{gather*}
The right hand side is
\begin{gather*}
         F_m(a_1)F_m(a_2)\Bigr(\sum_{i=1}^m  \frac{1}{F_m'(x_i)}
             \left[\frac{\partial}{\partial x}\left(
       \frac{f(x)(a_2-a_1)}{(x - a_1)^2(x-a_2)^2
               ( Q^{(2)}_m(x))^2F_m'(x) }\right)
           \right]_{x = x_i}
\end{gather*}
$$
       -  \sum_{k,l, k\neq l}
     \frac{2y_k y_l(a_2-a_1)}
{F_m'(x_k)F_m'(x_l)(x_l-a_1)(x_l-a_2)Q^{(2)}_m(x_l)(x_k-a_1)(x_k-a_2)
             Q^{(2)}_m(x_k)(x_l-x_k)}\Bigr).
$$
Then the proof of Theorem 3.1 is completely done due to next lemma.
\end{proof}

\begin{lemma}

\begin{enumerate}

\item
\begin{gather*}
\split
&\sum_{i=1}^m  \frac{1}{F_m'(x_i)}
             \left[\frac{\partial}{\partial x}\left(
       \frac{f(x)}{(x - a_1)^2(x-a_2)^2 (Q^{(2)}_m(x)^2
            F_m'(x) }\right) \right]_{x = x_i}\\
      &= \frac{2}{(a_1-a_2)^2}
\Bigr(\frac{f'(a_1)}{F_m(a_1)^2(Q^{(2)}_m(a_1))^2}
+\frac{f'(a_2)}{F_m(a_2)^2(Q^{(2)}_m(a_1))^2}\\
&+\frac{f'(a_{m+1})(a_1-a_2)^2}{
(a_{m+1}-a_1)(a_{m+1}-a_2)F_m(a_{m+1})^2 (Q^{(2)\prime}_m(a_{m+1}))^2}\\
&+\cdots \\
&+\frac{f'(a_{g})(a_1-a_2)^2}{
(a_g-a_1)(a_g-a_2)F_m(a_{g})^2 (Q^{(2)\prime}_m(a_g))^2}  \Bigr)    .\\
\endsplit
\end{gather*}

\item
$$
      \sum_{k,l, k\neq l}
     \frac{2y_k y_l(a_2-a_1)}
{F_m'(x_k)F_m'(x_l)(x_l-a_1)(x_l-a_2)Q^{(2)}_m(x_l)(x_k-a_1)(x_k-a_2)
             Q^{(2)}_m(x_k)(x_l-x_k)}\Bigr)
  =0
         .
$$
\end{enumerate}

\end{lemma}

\begin{proof}:
(1) will be proved by the following residual computations:
Let $\partial C_g^o$ be the boundary of a polygon representation
$C_g^o$ of $C_g$,
\begin{gather}
  \oint_{\partial  C_g^o} \frac{f(x)}
{(x-a_1)^2(x-a_2)^2F_m(x)^2(Q^{(2)}_m(x))^2} d x =0
 .           \label{eq:3-4}
\end{gather}
The divisor of the integrand of (\ref{eq:3-4}) is
$$
\sum_{i=1}^{2g+1}
(b_i,0) -4\sum_{i=1,2,m+1,m+2,\cdots,g}(a_i,0)-
 2\sum_{i=1}^m (x_i,y_i) - 2\sum_{i=1}^m (x_i,-y_i)
        +3\infty
$$
We check these poles:
First we consider the contribution around $\infty$ point.
\begin{gather*}
\split
\res_{(x_k, \pm y_k)}&
\frac{f(x)}{(x-a_1)^2(x-a_2)^2F_m(x)^2(Q^{(2)}_m(x))^2} d x\\
     & =  \frac{1}{F_m'(x_k)}
             \left[\frac{\partial}{\partial x}\left(
       \frac{f(x)}{(x - a_1)^2(x-a_2)^2 (Q^{(2)}_m(x))^2
       F_m'(x) }\right) \right]_{x = x_k}
        .\\
\endsplit
\end{gather*}
At the point $(a_1,0)$, noting that the local parameter
$t$ is given by $t = \sqrt{(x-a_1)}$ there, we have
$$
\res_{(a_1,0)}\frac{f(x)}{(x-a_1)^2(x-a_2)^2F_m(x)^2(Q^{(2)}_m(x))^2} dx
      = \frac{2f'(a_1)}{(a_1-a_2)^2F_m(a_1)^2(Q^{(2)}_m(a_1))^2}.
$$
The residue at $(a_2,0)$ is similarly obtained. For the points
$(a_k,0)$ $(g\ge k>m)$, we have
$$
\res_{(a_k,0)}\frac{f(x)}{(x-a_1)^2(x-a_2)^2F_m(x)^2(Q^{(2)}_m(x))^2} dx
      = \frac{2f'(a_k)}{(a_k-a_1)^2(a_k-a_2)^2
     F_m(a_2)^2(Q^{(2)\prime}_m(a_k))^2}.
$$
By arranging them, we obtain (1). (2) is obvious.
\end{proof}

\vskip 1.0 cm

As a corollary, we have Weierstrass relation (\ref{eq:1-4}) which was proved in
\cite{Ma}:

\begin{corollary}
For $m=g$ case, we have the Weierstrass relation for
a general curve $C_g$,
\begin{gather}
\frac{\partial}{\partial v_{r}^{(g)} }
          \frac{\partial}{\partial v_s^{(g)}}
      \log \frac{\al_r^{(g)}}{\al_s^{(g)}}
   = \frac{1}{(a_r-a_s)}
  \left(f'(a_r)
       \left(\frac{\al_s^{(m)}}{\al_r^{(m)}}\right)^2
+f'(a_s)
 \left(\frac{\al_r^{(m)}}{\al_s^{(m)}}\right)^2\right).
      \label{eq:3-5}
\end{gather}

\end{corollary}

Now we will give our finial proposition as  corollary.

\begin{corollary}
For a curve satisfying the relations
$c_j=a_j$ for $j=m+1,\cdots,g$,  $\al_r^{(m)}$ and  $\al_s^{(m)}$ $(r, s \in \{1,2,\cdots,m\})$
over $\Xi_m$ in (\ref{eq:2-5}) obey the relation,
\begin{gather}
\frac{\partial}{\partial v_{r}^{(m)} }
          \frac{\partial}{\partial v_s^{(m)}}
      \log \frac{\al_r^{(m)}}{\al_s^{(m)}}
   = \frac{1}{(a_r-a_s)}
  \left(\frac{f'(a_r)}{(Q^{(2)}_m(a_r))^2}
       \left(\frac{\al_s^{(m)}}{\al_r^{(m)}}\right)^2
+\frac{f'(a_s)}{(Q^{(2)}_m(a_s))^2}
 \left(\frac{\al_r^{(m)}}{\al_s^{(m)}}\right)^2\right).
      \label{eq:3-6}
\end{gather}
\end{corollary}

\begin{proof} Since the condition $c_j=a_j$ for $j=m+1,\cdots,g$
means $f'(a_j)=0$ for $j=m+1,\cdots,g$, Theorem 3.1 reduces to
this one.
\end{proof}

Under the same assumption of Corollary 3.4,
letting
$
A=\dfrac{2\sqrt{f'(a_r)f'(a_s)}}{(a_r-a_s)Q_m(a_r)Q_m(a_s)},
$ and
$$
   \phi_m^{(r,s)}(u) :=\frac{1}{\sqrt{-1}} \log
          \sqrt{\frac{f'(a_r)}{f'(a_s}}
         \frac{Q_m(a_r)}{Q_m(a_s)} \frac{F_m(a_r)}{F_m(a_s)},
$$
defined over $\Xi_m$, $\phi_m^{(r,s)}$ obeys the sin-Gordon equation,
\begin{gather}
\frac{\partial}{\partial v_{r}^{(m)} }
          \frac{\partial}{\partial v_s^{(m)}}
     \phi_m^{(r,s)}=A\cos( \phi_m^{(r,s)}).
      \label{eq:3-7}
\end{gather}

\vskip 0.5 cm

\

\end{document}